# Transport Properties of Topological Insulators: Band Bending, Bulk Metal-to-Insulator Transition, and Weak Anti-Localization

Matthew Brahlek[1], Nikesh Koirala[1], Namrata Bansal[2], and Seongshik Oh[1,*]

[1]Department of Physics and Astronomy, Rutgers, The State University of New Jersey, Piscataway, New Jersey 08854, USA

[2]Department of Electrical and Computer Engineering, Rutgers, The State University of New Jersey, Piscataway, New Jersey 08854, USA

*Corresponding author: ohsean@physics.rutgers.edu

**ABSTRACT:** **We reanalyze some of the critical transport experiments and provide a coherent understanding of the current generation of topological insulators (TIs). Currently TI transport studies abound with widely varying claims of the surface and bulk states, often times contradicting each other, and a proper understanding of TI transport properties is lacking. According to the simple criteria given by Mott and Ioffe-Regel, even the best TIs are not true insulators in the Mott sense, and at best, are weakly-insulating bad metals. However, band-bending effects contribute significantly to the TI transport properties including Shubnikov de-Haas oscillations, and we show that utilization of this band-bending effect can lead to a Mott insulating bulk state in the thin regime. In addition, by reconsidering previous results on the weak anti-localization (WAL) effect with additional new data, we correct a misunderstanding in the literature and generate a coherent picture of the WAL effect in TIs.**

## 1. Introduction

Topological insulators (TIs) are a new class of matter that do not follow Landau's classification by spontaneously broken symmetry; rather they are distinguished only by considering which topological class they belong. The topological invariant is derived from the band structure properties of the bulk electrons, and in certain materials, sufficiently strong spin-orbit coupling causes the valence and conduction bands to invert, which results in the topological invariant to change and a topological surface states (TSS) to develop (see Ref. [1-4] and references therein). In the case of three dimensional (3D) TIs, such as $Bi_2Se_3$, and $Bi_2Te_3$, it was predicted and observed that the surface states span the bulk energy gap and have a linear Diraclike dispersion with chiral spin-momentum locking[5-7]. Unlike typical surface states in insulators and metals, the surface states in TIs have an odd number of Dirac points, which are protected by time-reversal symmetry against localization effects. However, it was soon realized that all the materials that have these Dirac-like TSS are in fact not true TIs; all the known materials suffer from significant charge defects that push the Fermi level ($E_F$) into either the conduction band (CB) or the valence band (VB), and therefore should be referred to not as TIs, but rather as topological conductors. This parallel channel complicates transport experiments because it is difficult to distinguish between effects from the TSS and those that originate from this bulk state.

Here we present a coherent picture for understanding transport measurements on the current generation of TIs. In section II we develop and apply the simple Mott and the Ioffe-Regel criteria for metal-to-insulator transitions and show that the majority of the best TI samples are in fact still metallic in the bulk. In section III, we develop and discuss band-bending effects, which helps understand some of the widely varying data in the literature particularly focusing on Shubnikov de-Haas oscillations. Then, we show that the band-bending effect can help overcome the Mott limit and a truly bulk insulating state can develop in the thin regime. Finally, in section IV we discuss the various aspects of the weak anti-localization (WAL) effect and their thickness dependences in TIs. This is all done in the context of experimental data.

## 2. Bulk Metal-to-Insulator Transition

A condition for a material to become topologically non-trivial is for certain energy bands to invert (other promising examples such as topological Kondo insulators have recently been predicted[8] with experiments in the initial phase[9-13]). Band inversion is most easily achieved in narrow gap semiconductors that are predicted to have non-inverted bands, and by incorporating realistic spin-orbit coupling the energy bands invert[7]. However, in real materials there are a finite number of thermodynamically induced atomic vacancies and anti-site defects that occur and act as charge defects, which shifts $E_F$. Since the band gap is narrow, $E_F$ is easily moved into the bulk energy bands, and the material becomes metallic in the bulk. To get a feeling for how the bulk $E_F$ changes as a function of bulk charge defect density, $N_{BD}$, we have plotted $E_F = \hbar^2/(2m^*)(3\pi^2 N_{BD})^{2/3}$ in Fig. 1, where we assumed an isotropic Fermi surface with the effective mass given in terms of the electron mass to be $m^* \approx 0.15 m_e$, a typical value for $Bi_2Se_3$ and $Bi_2Te_3$[14]. Here and for all the following discussions, unless explicitly specified, the temperature is assumed to be absolute zero, which is a valid assumption for the commonly measured transport properties of TIs. What can be seen from this is that $E_F$ is pinned within a few meV of the conduction band minimum (or valence band maximum in p-type materials) when the carrier density drops below $\sim 10^{17}$-$10^{18}$/cm$^3$. $E_F$ can only become unpinned from the conduction band minimum and fall into the band gap if a metal-to-insulator transition occurs with reduced carrier densities; these transitions can occur via two different processes, Mott transition and Ioffe-Regel criterion.

When a dopant atom is embedded in an insulator, it forms a rescaled Coulomb potential, and this leads to hydrogenlike bound states, each with an effective Bohr radius given by $a_B = \varepsilon(m_e/m^*) \times 0.5$ Å, where $\varepsilon$ is the dielectric constant, $m_e$ is the electron mass, $m^*$ is the effective mass, and 0.5 Å is the free space Bohr radius. In a crystal, as the dopant density is increased, the mean separation of the dopant atoms decreases and eventually becomes comparable to the effective Bohr radius. When this happens, electrons that were bound to the dopant sites can become delocalized and freely move to neighboring sites, which gives rise to the insulator becoming a metal. The Mott criterion predicts that this happens when the dopant density passes the critical density given by $a_B N_M^{1/3} \approx 0.26$ (see Ref. [15]). Rearranging this, we then get the critical value to be $N_M \approx (m^*/(\varepsilon m_e))^3 \times (1.4 \times 10^{23} /cm^3) \approx 3 \times 10^{14} /cm^3$ for both Bi$_2$Se$_3$ and Bi$_2$Te$_3$; this was found by using $\varepsilon \approx 110$ and $m^* \approx 0.15 m_e$ [14]. $N_M \approx 3 \times 10^{14} /cm^3$ is extremely small compared to the corresponding numbers for common semiconductors such as Si where $N_M \approx 10^{18} /cm^3$. In fact, for Bi$_2$Se$_3$ the lowest dopant density achieved thus far is $\sim 10^{16} /cm^3$ (see Ref. [14]); it was found that these crystals were dominated by bulk carriers, which can be rationalized by considering the Mott criterion. In standard semiconductors such as Si, the bonding is very strong, and therefore a defect density of $\sim 10^{14} /cm^3$ is achievable in the cleanest materials. However in current TI materials, the bonding is much weaker, and therefore such a low defect density may be thermodynamically impossible to achieve either in intrinsic or in compensation-doped materials.

The Ioffe-Regel criterion explains how a metal transitions into an insulator with increased disorder and reduced carrier density. It states that the electrons in a metal become localized when the mean free path drops below the Fermi wave length[16]. In other words, the system remains metallic only if $k_F l > 1$, where $k_F$ is the Fermi wave vector and $l$ is the mean free path, which can easily be calculated based on typical values measured by transport. Assuming a 3D isotropic Fermi surface $k_F = (3\pi^2 N_{BD})^{1/3}$ and $l = (\hbar\mu/e)(3\pi^2 N_{BD})^{1/3}$ where $\mu$ is the electron mobility, $e$ is the electron charge, and $N_{BD}$ is the volume carrier density, which is related to the Hall carrier density by $N_{BD} = n_{Hall}/t$ with $t$ being the crystal's thickness. By setting $k_F l = 1$, this can be rearranged into a function of critical mobility, $\mu_{IR}$, vs $N_{BD}$; the result is $\mu_{IR} = (e/\hbar)(3\pi^2 N_{BD})^{-2/3}$. What this says is that if $\mu < \mu_{IR}$ then the system can transition into a weakly insulating phase even if the bulk defect density exceeds the critical value predicted by the Mott criterion. Such an insulating phase usually can be described by a variable range hopping mechanism, in which the conduction electrons typically hop a certain distance before becoming localized[17]. However, these types of insulating phases tend to remain quite conducting at any finite temperatures, only truly insulating in the limit of zero temperature, and are thus often called a bad metal.

The results for the Mott and the Ioffe-Regel criteria are summarized in Fig. 2, where data have been compiled from single crystals and thin films of Bi$_2$Se$_3$, Bi$_2$Te$_3$, Bi$_2$Te$_2$Se as well as these compounds that have been intentionally charge-doped to compensate for the native carriers[14, 18-29]. In Fig. 2, the vertical line corresponds to the Mott criterion, where everything to the left is in a strong insulating state. The diagonal region roughly corresponds to the Ioffe-Regel criterion of $k_Fl$ ~1. The region below this, but to the right of the Mott criterion, is in a bad metal phase. The region above the Ioffe-Regel criterion and to the right of the Mott criterion is a good metal. What is clear from this plot is that no samples come close to the critical density predicted by the Mott criterion. Currently, the samples that are closest to insulating are $Bi_2Te_2Se$ bulk crystals, which show insulatorlike behavior in resistance vs temperature but have significant bulk carriers and are highly disordered with extremely small mobilities. These samples become more insulating at the cost of decreasing mobility. Optimally, one would want to move left horizontally, while maintaining the mobility.

Last of all, the Mott criterion gives critical insight in the position of the Fermi level deep inside the bulk. Since all the transport data give a bulk carrier density that exceeds the Mott criterion by at least a few orders of magnitude, it is then implied that $E_F$ deep in the bulk must be at the conduction band minimum. This does not contradict the Ioffe-Regel criterion which says the bulk electrons can be localized even if $E_F$ is at the conduction band minimum, simply by making the system more disordered. We will see next that this is an important point when band-bending effects are incorporated.

**3. Band-bending effects**

Band bending occurs wherever there is a mismatch in Fermi levels, for example, at the interface between two semiconductors. In any case, for band bending to occur there needs to be charge transfer to equilibrate the Fermi level everywhere. This charge transfer then creates an electric field, and the associated potential shifts the bands in the vicinity of the charge transfer. In $Bi_2Se_3$ the band bending is due to a large mismatch between the surface and bulk Fermi levels, and then upon equilibrating, the bulk band shifts near the surfaces. This shift has been found in some samples to be upward giving rise to a depletion of carriers near the interface[24-25] [30], while in other samples it has been found to shift downward, which gives rise to an accumulation region near the surface[27, 31]. This downward bending also gives rise to a non-topological two dimensional electron gas (2DEG) because the downward bending creates a quantum well that quantizes the bulk bands near the surface[31-37].

The band bending direction is strongly dependent on both the surface and bulk defect levels. Given that the Mott criterion fixes $E_F$ to be at the conduction band minimum and knowing the surface state carrier density, together these fix the band bending as follows[30]. First, assume that at the surface $E_F$ barely touches the CB minimum. Then assuming a Fermi velocity ($v_F \approx 4.0 \times 10^5$ m/s) and the energy of the CB minimum relative to the Dirac point ($E_{DP} \approx 210$ meV [31, 38]), the surface carrier density can be calculated for one surface to be $n_{SS} \approx 5.0 \times 10^{12}$ /cm$^2$. Since $E_F$ at the surface and in the bulk is at the CB minimum, the bands must be flat, and band bending is negligible (see Fig. 3(a)). Therefore, $n_{SS} \approx 5.0 \times 10^{12}$ /cm$^2$ gives the flat band condition. Now, if $n_{SS}$ increases above ~$5.0 \times 10^{12}$ /cm$^2$ then the conduction band bends downward near the surface, giving rise to an accumulation of carriers near the surface (see Fig. 3(b)). Alternatively, if $n_{SS}$ is less than ~$5.0 \times 10^{12}$ /cm$^2$ then the bands bend upward close to the surface, which implies that there should be a depletion of bulk carriers near the surface (see Fig. 3(c)).

There is a large difference between upward and downward band bending. In the case of upward band bending with bulk carriers less than ~$10^{18}$ /cm$^3$, the Poisson equation is sufficient to quantify the effect (see below). In the case of downward band bending, the Thomas-Fermi approximation is valid because the system near the surface is much more like a typical metal. For Thomas-Fermi screening, the length scale over which the screening occurs is given by $l_0 \approx ((\varepsilon_0 \pi^2 \hbar^2)/(k_F m^* e^2))^{1/2} \approx 6$ Å, where $k_F \approx 0.07$ /Å is the Fermi wave vector as typically measured by ARPES, $m^* \approx 0.15 m_e$ is the effective mass in terms of the electron mass $m_e$, $e$ is the electron charge, and $\varepsilon_0$ is the free space dielectric constant. This shows that for downward band bending the screening length is ≲ 1 nm. In contrast, for the case of upward band bending, the length scale is much larger, as will be shown below. Although downward band bending is more common with many TI materials, we will show below that if upward band bending can be achieved in thin film TIs, truly insulating bulk state could be implemented even in current generation TIs, whose bulk states are always metallic under the Mott criterion.

Page **6** of **30**

In general, band bending should be treated in a self consistent manner with a numerical solution to the coupled Poisson-Schrodinger equations. However, the case of upward band bending, shown in Fig. 3(c), can be described to first order by considering only the Poisson equation. This formalism is only valid in the limit of when $E_F$ is close to the CB minimum and when the bulk defect density, $N_{BD}$, is uniform. From Fig. 2, we can see that $E_F$ begins to deviate from the CB bottom significantly when $N_{BD}$ is larger than ~$10^{18}$/cm$^3$, and therefore, in the following we assume $N_{BD} < 10^{18}$/cm$^3$; high quality TI samples (bulk crystals and thin films) with minimal defect densities satisfy this condition. The Poisson equation is

$$\nabla^2 V(z) = -\frac{e^2 N_{BD}}{\varepsilon \varepsilon_0} \quad (1)$$

where $V(z)$ is the potential as a function of $z$, the depth from the surface of the TI. Then, by assuming $N_{BD}$ is roughly constant and the potential is fully screened at a depth of $z_D$, the depletion depth, the solution for one surface is given by

$$V(z) = \begin{cases} \frac{e^2 N_{BD}}{2\varepsilon\varepsilon_0}(z-z_D)^2 & z < z_D \\ 0 & z > z_D \end{cases}. \quad (2)$$

At $z = 0$, the CB minimum is at its largest value (see Fig 4(a)), $V(z = 0) = \Delta V$, which gives the following equation

$$\Delta V = \frac{e^2 z_D^2 N_{BD}}{2\varepsilon\varepsilon_0}. \quad (3)$$

This is summarized in Fig. 4(a). The band bending can be related back to the various charge densities by the following. There are two intrinsic defect levels, the bulk defect density, $N_{BD}$, and the surface defect density, $n_{SD}$, which together set the initial surface and bulk Fermi levels, $E_{F,SD}$ and $E_{F,BD}$ (see Fig. 4(b)) prior to equilibrium being established. For equilibrium to then be established, the Fermi level on the surface needs to match that of the bulk, and therefore a certain amount of charge, $n_{CT}$, needs to be transferred from the bulk onto the surface (see Fig. 4(c)). Since we assumed a constant bulk defect density, $n_{CT}$ is simply given by

(4)
$$n_{CT} = N_{BD} \times z_D.$$

So now, the total amount of charge on the surface is

(5)
$$n_{SS} = n_{CT} + n_{SD}.$$

Lastly, we can get one more equation by looking at Fig. 4(a). By assuming $E_F$ is at the conduction band bottom deep inside the bulk, we can see that the surface Fermi level, relative to the Dirac point, is related to the level of band bending, $\Delta V$, by

(6)
$$E_{F,SS} = \hbar v_F \sqrt{4\pi n_{SS}} = E_{DP} - \Delta V,$$

where we used the dispersion relation for a Dirac state $E_{F,SS} = \hbar v_F k_F = \hbar v_F (4\pi n_{SS})^{1/2}$, where $n_{SS} = k_F^2/(4\pi)$, and $E_{DP} \approx 210$ meV is the separation between the conduction band minimum and the Dirac point. Combining Eq. (3 - 6) gives a set of 4 equations with 6 unknowns, which allows us to solve the system by letting $\Delta V$, $z_D$, $n_{SS}$, and $n_{CT}$ be the dependent variables, which depend on $n_{SD}$ and $N_{BD}$. We proceed this way because experimentally $n_{SD}$ and $N_{BD}$ are set by growth conditions and the subsequent exposure of the surface to the experimental environment (vacuum, atmosphere, etc). An analytic solution to this set of equations exists. However, it is too complicated to present explicitly, and the clearest way to understand the physics of the solution is the contour plots shown in Figs. 4(d)-4(g) as a function of $n_{SD}$ and $N_{BD}$.

The contour plots in Fig. 4(d)-4(g) contain a large amount of information and can be understood by the following. The vertical axes are the intrinsic defect density ($n_{SD}$) on the surface, while the horizontal axes are the intrinsic bulk defect density ($N_{BD}$). In Fig. 4(d) $\Delta V$ is plotted. For fixed $N_{BD}$, as $n_{SD}$ increases $\Delta V$ decreases accordingly because the $n_{SD}$ sets the initial $E_F$ on the surface, and the closer the surface $E_F$ is to the bulk conduction band minimum, the less the bands need to bend. In Fig. 4(f) $z_D$ is plotted. As $N_{BD}$ decreases, $z_D$ increases, which is expected because it takes a larger distance of the depletion zone to raise the surface $E_F$ to that of the bulk state; in a similar way $z_D$ decreases for increasing $n_{SD}$. Figures 4(e) and 4(g) show the total surface states carrier density, $n_{SS}$, and the total charge transfer, $n_{CT}$. This shows that for low bulk carrier density ($\lesssim 5\times 10^{16}$ /cm$^3$) the carrier density on the surface is dominated by the surface defect level, and only when the bulk carrier density exceeds ~$5 \times 10^{16}$ /cm$^3$, is there significant charge transfer to the surface.

Figures 4(d)-4(g) allows us to understand what happens when Bi$_2$Se$_3$ and Bi$_2$Te$_3$ are exposed to atmosphere or controlled surface contamination in ARPES experiments[25, 37]. The long term effect of atmospheric exposure has been shown to increase the carrier density[39]. What can be seen from Fig. 4 is that if the crystals start with upward band bending and low carrier densities on the surface and in the bulk, then as time increases the band bending flattens out, and upward band bending will eventually give way to downward bending; such an effect has been linked to charge doping by water vapor in atmospheric gases[24, 39-41]. This change of band bending character will be accompanied with the development of quantum well states on the surfaces, which are in parallel with the TSS.

The level and direction of band bending is crucial to observing Shubnikov de-Haas (SdH) oscillations from the TSS. There has been a puzzling difference between two sets of Bi$_2$Se$_3$ based samples reported in Analytis *et al*[24] and Butch *et al*[14]. In the former study, SdH oscillations observed at high magnetic field were consistent with the TSS, while in the latter study no surface SdH oscillations were observed. The puzzle is that Butch *et al*'s samples were an order of magnitude lower in bulk defect density and still did not show surface SdH oscillations. The main difference between the samples in the two studies was the addition of a small amount of Sb in the Analytis *et al*'s. The addition of the Sb makes the initial state of the samples start with upward band bending[25], while pure Bi$_2$Se$_3$ starts in an initial state with downward band bending. As shown in Figs. 3(b) and 3(c), if the band bending is downward, the size of the Fermi surface of the TSS is much larger than if the bands bend upward. This affects the SdH oscillations from the TSS in two ways. First, the period of SdH oscillations depends inversely on the cross sectional area of the Fermi surface; consequently a large Fermi surface translates into a small period of oscillation, and thus it is harder to observe the SdH oscillations from the surfaces with downward band bending. The second complication is that with downward band bending, quantum well states, which are bulk states confined near the surface, develop in the conduction band[37]. These states can have higher mobility than the topological surface states and since their Fermi surfaces are smaller than the TSS, the carrier density tends to be lower than that of the TSS. These together with the natural 2D nature of the quantum well states can make their SdH oscillations more prominent and easily confusable with the SdH oscillations that emanate from the TSS.

Since the amplitude of the SdH oscillations is exponentially sensitive to the mobility, the mobility dependence on band bending is another factor affecting observation of the SdH oscillations from the TSS. Band bending can affect the mobility from the TSS in two ways, both of which can be seen by considering the Drude mobility, $\mu = e\tau/m^*$, where $\tau$ is the carrier relaxation time and $m^*$ is the effective mass (more specifically $m^*$ is the cyclotron mass for a Dirac band). The first is that $\tau$ is related to the change in band bending direction. As mentioned above, for upward band bending to be present, the surface defect density needs to be sufficiently low, and if the surface defect density is too large then the band bending direction is downward. Because more defects imply shorter relaxation times, this directly implies that $\tau_{upward} > \tau_{downward}$. Secondly, for a Dirac band $m^*$ is proportional to the Fermi level measured from the Dirac point, $m^* = E_F/v_F^2$, whereas for a parabolic bulk band $m^*$ is constant. Because the Fermi level is lower for upward band bending than for downward bending while the Fermi velocity remains almost constant, this then implies $m^*_{upward} < m^*_{downward}$. Both effects lead to $\mu_{upward} > \mu_{downward}$. Altogether, this helps explain why Butch *et al*, whose samples had downward band bending, saw no sign of SdH oscillations from the TSS while those samples measured by Analytist *et al* that exhibited upward band bending revealed SdH oscillations consistent with the TSS.

Finally, from this analysis of upward band bending we can get some insight into the regime where the thickness is comparable with the depletion length. For the above analysis, it was assumed that the thickness of the crystal was much larger than the depletion length as shown in Fig 5(a). On the other hand, when the thickness drops below twice the infinite limit depletion length, the solution to the Poisson equation (Eq. (2)) becomes

(7) $$V(z) = \frac{e^2 N_{BD}}{2\varepsilon\varepsilon_0}(z - t/2)^2 \ (0 < z < t),$$

where $t$ is the thickness of the crystal. Since in this finite thickness limit $t < 2 \times z_D$, the charge transferred from the bulk state to the surface is now limited to

(8) $$n_{CT,F} = N_{BD} \times t/2 < N_{BD} \times z_D.$$

This implies then that the bulk is now completely depleted because all the free electrons are transferred out of the bulk onto the surface. As mentioned above, before equilibrium is established, the Fermi levels on the surface and in the bulk are set by the defect levels that reside in those regions, and the charge transfer raises the surface Fermi level up to that of the bulk. Now, if the available charge in the bulk, $N_{BD} \times t/2$, is insufficient to raise the Fermi level on the surface to the bottom of the conduction band, then all the free electrons are transferred out of the bulk and the bulk Fermi level simply drops below the conduction band minimum to establish equilibrium. In fact, this is not in contradiction with the Mott criterion, because the Mott criterion states that the Fermi level must be pinned in the conduction band only until the bulk carrier density remains above $N_M$ (~$3 \times 10^{14}$ /cm³) as shown in section II.

Following the analysis given above we can actually calculate the gap, $\Delta$, between the conduction band minimum and the Fermi level as a function of film thickness, $t$. The main modification is to Eq. (6), which becomes

(9) $$E_{F,SS} = \hbar v_F \sqrt{4\pi n_{ss}} = E_{DP} - \Delta V - \Delta(t),$$

where the Fermi level has settled at $\Delta(t)$ below the conduction band minimum (see Fig. 5(c)). As was done with Eq. (6), Eq. (9) can be expanded in terms of the thickness and the bulk and surface defect densities, and then we can solve for $\Delta(t)$ only in terms of these parameters. This procedure yields

(10) $$\Delta(t) = E_{DP} - \frac{e^2 N_{BD}}{2\varepsilon\varepsilon_0}\left(\frac{t}{2}\right)^2 - \hbar v_F\sqrt{4\pi(n_{SD} - N_{BD}t/2)},$$

where, in the finite thickness limit, we used $\Delta V = (e^2 N_{BD}/(2\varepsilon\varepsilon_0))(t/2)^2$ and $n_{SS} = n_{SD} + n_{CT,F} = n_{SD} + N_{BD}(t/2)$. Eq. (10) shows that if the bulk and surface defect levels are low enough, there exists a critical thickness for which the Fermi level drops below the conduction band minimum. This can be easily seen by plotting Eq. (10) for fixed defect densities against thickness in Fig. 5(d) or for fixed thickness in the contour plots in Fig. 5(e)-(g). Figure 5(e) shows that in the ultra thin limit ($t < 10$ nm) the Fermi level should be below the conduction band minimum, so long as the surface carrier density is below the critical value where upward band bending gives way to downward band bending (as mentioned previously, this occurs when $n_{SS} > 5.0 \times 10^{12}$ /cm$^2$ for one surface). In contrast, in the bulk limit ($t > 1000$ nm) in Fig. 5(f), $\Delta(t)$ remains zero for experimentally achievable bulk defect density levels, which are $\sim 10^{16}$/cm$^3$ or higher, implying that Fermi levels deep inside the bulk still remain pinned to the bottom of the conduction band in bulk crystals. Altogether, this suggests that even if the bulk charge defect density is higher than the Mott criterion of $\sim 3\times 10^{14}$ /cm$^3$, as far as the surface charge density is kept below the critical value ($\sim 5\times 10^{12}$ /cm$^2$) necessary for the upward band bending, making the samples thin will force the bulk Fermi level to drop below the conduction band minimum and lead to a truly insulating bulk state. In other words, through thin film engineering, it should be possible to achieve Mott insulating bulk state even in current generation TIs. The main experimental obstacle is that most TI materials such as $Bi_2Se_3$ have strong tendency toward downward, instead of upward, band bending. However, it was recently demonstrated that Cu doping in $Bi_2Se_3$ thin films can help stabilize the upward band bending and it does lead to the bulk insulating state as expected here[30].

### 4. Weak Anti-Localization

Weak anti-localization (WAL) and weak localization (WL) are effects in transport measurements where the conductance is enhanced or suppressed due to coherent back scattering, and are usually seen as a sharp change in the conductance with the application of a small magnetic field. The origin of this effect is due to coherent back scattering of the electrons along time reversible paths (see Ref. [42] for more details). As shown in Fig. 6, an electron can trace out many closed paths, which return the electron to its original path with opposite direction. The paths that contribute to this effect are those where no inelastic scattering events occur; such events cause the electrons to lose their phase information, which suppresses the interference effect. The amplitude to scatter backwards is then the sum of the amplitudes to go around the path clockwise (CW) and counter clockwise (CCW), during which the electron accumulates a phase that is given by the spin operator for an electron, $S$, dotted into the angle operator, $\theta = \phi n$, where $\phi$ is the angle accumulated in the spin part of the wave function and $n$ is the unit vector along the rotation direction. This then yields

$$\psi_{CW} = \psi_0 e^{-i\frac{S}{\hbar}\cdot\theta_{CW}} = \psi_0 e^{\frac{i\phi}{2}}$$

$$\psi_{CCW} = \psi_0 e^{-i\frac{S}{\hbar}\cdot\theta_{CCW}} = \psi_0 e^{\frac{-i\phi}{2}}$$

and the amplitude $\psi_0$ contains any other phase information accumulated in the spatial part of the wave function; in the absence of magnetic field, magnetic impurities or other dephasing scattering events, these two time reversal paths share the same spatial amplitude of $\psi_0$. The amplitude to return to the original position is then given by the sum of the CW and CCW components

$$\psi = \psi_0 e^{\frac{i\phi}{2}} + \psi_0 e^{\frac{-i\phi}{2}} = 2\psi_0 \cos\left(\frac{\phi}{2}\right).$$

From this we can see that there is a big difference between electrons without spin-momentum locking and those with strong spin momentum locking. For the former case, the angle the electron's spin sweeps out as it traverses the paths is negligibly small so $\psi_{WL} \approx 2\psi_0$. Compared with the case of no interference (i.e. classical case), which yields a probability of $2|\psi_0|^2$ for backscattering, this constructive interference doubles the probability of backscattering to $4|\psi_0|^2$. This enhanced backscattering effect is the root of the weak localization effect. For the latter case, where there is significant spin-momentum locking, the angle the spin sweeps out as it traverses the time reversible paths is $\pi$ so $\psi_{WAL} \approx 2\psi_0\cos(\pi/2) \approx 0$. This destructive interference effect suppresses the backscattering leading to the weak anti-localization effect.

Now, an applied magnetic field breaks the time reversal symmetry and the two opposite paths accrue opposite phases of equal magnitude, $\beta = e\Phi/\hbar$, which is proportional to the magnetic flux, $\Phi$, through the time reversible loops. With this additional phase factor the wave function becomes

$$\psi = \psi_0 e^{i\beta} e^{\frac{i\phi}{2}} + \psi_0 e^{-i\beta} e^{\frac{-i\phi}{2}} = 2\psi_0 \cos(\frac{\phi}{2} + \beta)$$

and the exact destructive (WAL) or constructive (WL) interference becomes dephased as the magnetic field (hence $\beta$) increases. This phase factor strongly depends on the flux through the loops, and therefore with a small increase in the magnetic field, the localization effect quickly dies out: however, the periodic dependence on the magnetic field as implied by this formula does not actually occur because the area that determines the flux varies with different electron paths, and the correct treatment requires more rigorous approaches such as the HLN formalism to be discussed below. This gives rise to the sharp cusp at low magnetic field, which is a signature of WAL and WL as shown in Fig. 6. In WAL (WL) the destructive (constructive) interference effect suppresses (enhances) the backscattering and leads to enhanced (reduced) conductance in the absence of magnetic field in Fig. 6(b) (Fig. 6(a)). The magnitude of this enhancement (reduction) of the conductance due to WAL (WL) is typically of the order of 10% of the overall conductance in TIs, which is significant, yet implying that the other scattering channels account for the majority of the scattering.

The WAL and WL effects are typically analyzed using the HLN formalism, which was developed in the 1980s for 2D systems[43]. The following formula was derived to quantify the correction to the conductance: $\Delta G(B) = \tilde{A} e^2/(2\pi h)[ln(B_\phi/B) - \Psi(1/2 + B_\phi/B)]$, where $h$ is Planck's constant, $\tilde{A}$ is a parameter related to the number of 2D conductive channels (1 for each channel), $B_\phi$ is the de-phasing field, and $\Psi(x)$ is the digamma function. The two independent parameters are $\tilde{A}$ and $B_\phi$ (in the original HLN formalism, the prefactor was in units of $1/(2\pi)$ per channel, but for simplicity, we incorporate this $2\pi$ term into the prefactor so that $\tilde{A} = 1$, instead of $1/(2\pi)$, implies a single channel).

$B_\phi$ is a magnetic field that roughly gives the width of the WAL cusp in the $\Delta G$ vs $B$ plot. To get a physical understanding, $B_\phi$ can be related to a length scale by $B_\phi = \hbar/(4el_\phi^2)$, where $l_\phi$ is the de-phasing length. Roughly speaking all time reversible paths that are within the de-phasing length will contribute to the WAL effect. Such a parameter can then be thought of like a mean-free path, which is dependent on the microscopic details such as disorder. In contrast, the other parameter, $\tilde{A}$, is less sensitive to such details; $\tilde{A}$ is related to the number of channels in the system. For an ideal TI with surface states on the top and bottom, each should carry a channel number of 1, and if $B_\phi$ is similar, then the conductance should add, which would give an effective $\tilde{A} = 1 + 1 = 2$. However, in all known TIs, there is a metallic bulk channel, which complicates the analysis of this parameter. Despite this, $\tilde{A}$ provides vital clues to help unravel transport measurements.

In thin-film samples of $Bi_2Se_3$ $\tilde{A}$ is not the ideal value of 2, but rather close to half of this at $\tilde{A} \approx 1$[27, 44-48] (see Fig. 7). This suggests that there is a conducting bulk state in parallel with the two topological surface states. Electrons on the surface can couple to this bulk state, and together the entire film acts effectively as one channel, and therefore $\tilde{A} \approx 1$; this has been confirmed by electro-static gating experiments[46, 49]. As the gate-voltage depleted the bulk state near one surface, $\tilde{A}$ gradually increased from ~1 towards ~2, implying that two surface states mediated by the conducting bulk behave as one 2D channel, whereas two isolated surface states appear as two 2D channels in the WAL effect.

This $\tilde{A}$ parameter is therefore a sensitive indicator to distinguish between samples that have conducting bulk versus those with the bulk state depleted. Two experiments have made the best case for a truly depleted bulk state, one by surface-doping thin $Bi_2Se_3$ flakes in conjunction with electro-static gating[50], and the other achieved with compensation-doped $Bi_2Se_3$ thin-films[30]. The likely explanation as to how these experiments got around the Mott criteria is the previously mentioned use of band bending in the ultra thin limit ($t < 2 \times z_D$). The main result of both experiments is the observation that the $\tilde{A}$ parameter transitions from 1 to 2 as the bulk state transforms from metallic to insulating state. The former experiment by D. Kim *et al*[50] showed that while $\tilde{A} \approx 1$ regardless of thickness when the carrier density is high ($\gtrsim 5 \times 10^{12}$ /cm$^2$ per surface), as the carriers were depleted through gating $\tilde{A}$ transitioned from 1 to 2 for films thicker than ~10 QL but remained at ~1 for thinner samples. The latter experiment by Brahlek *et al* [30] provided a more extensive thickness dependence of the WAL parameter in Cu-doped ultralow carrier density (~2.5 × 10$^{12}$ /cm$^2$ per surface) Bi$_2$Se$_3$ films. Like D. Kim *et al*, Brahlek *et al*, first showed that at fixed thickness (20 QL) as the bulk state was depleted by Cu doping, $\tilde{A}$ smoothly increased from 1 to 2. They further showed that when the bulk state is made insulating, for films below 10 QL, $\tilde{A}$ was exactly quantized at 1, but as the films were made thicker than 10 QL, the $\tilde{A}$ parameter increased from 1 towards 2, and from 20 QL to 150 QL $\tilde{A}$ remained quantized at 2: these data are reproduced in Fig. 7(b). Together, both studies suggest that when the coupling between the top and bottom surfaces is suppressed, the two surfaces contribute individually to the total channel number of two for $\tilde{A}$, but if coupling is allowed between the two surfaces either through bulk conduction or through tunneling in the ultrathin regime, they contribute as a whole one channel to $\tilde{A}$.

Lastly, WAL sometimes shows a transition from $\tilde{A} = 1$ to 0 in the ultrathin films [45, 47]. Considering that a hybridization gap is formed at the Dirac point when Bi$_2$Se$_3$ films get thinner than ~6 QL according to an ARPES study[51], such a transition is naturally expected when the Fermi level is tuned into this gap and was experimentally confirmed by D. Kim *et al*[50]. However, similar transitions were observed even in films with their Fermi levels far above the Dirac point: first by us in Y. Kim *et al*[45] and later by Taskin *et al*[47]. Although these studies report that the transition occurs at just below 5 QL, which happens to be close to the gap opening thickness as observed by the ARPES study, whether this is related to the gap opening at the Dirac point is questionable considering that their Fermi levels are far (~300-500 meV) from the Dirac point; disorder and surface scattering in the ultrathin regime can easily drive otherwise metallic films into insulating state causing such a 1-to-0 WAL transition. Although Y. Kim *et al* did not interpret it in terms of the gap opening process, Taskin *et al* interpreted a similar 1-to-0 transition as due to surface-hybridization gap opening at the Dirac point even though their surface

Fermi levels were far from the gap. If this argument were correct, such a transition should have been observed at similar thicknesses in $Bi_2Se_3$ films regardless of their Fermi levels.

However, Bansal et al[27], Liu et al[52], and Brahlek et al[30] showed that $Bi_2Se_3$ films with much higher mobilities did not exhibit such a 1-to-0 transition in $\tilde{A}$, which remained close to 1 all the way down to 2 QL. In Fig. 7, we compare the published thickness-dependent WAL channel data from $Bi_2Se_3$ films and a new data set on amorphous $SiO_2$: the mobilities at ~1.5 K for their 2, 3 or 6 QL samples are also shown for quantifying the level of disorder in the ultrathin regime. The comparison shows that there is no universality in the 1-to-0 WAL transition and it strongly depends on, among other things, the mobility of the samples. While the higher mobility samples by Bansal et al, Liu et al, or Brahlek et al did not exhibit any transition all the way down to 2 QL, the lower mobility samples by Y. Kim et al and Taskin et al exhibited the 1-to-0 transition between 3-5 QL, and those on amorphous $SiO_2$, which had the lowest mobilities in the thin regime among the compared results, showed the transition at an even higher thickness. Furthermore, all the lower mobility samples exhibited clearly insulating temperature dependence at low temperatures, whereas the higher mobility samples remained pretty much metallic down to the lowest measurement temperatures. This mobility dependent metal-to-insulator transition can be partly understood with the above-discussed Ioffe-Regel criterion in the 3D context. It is also well known that 2D metallic films turn insulating once their sheet resistance becomes similar to or larger than the quantum resistance ($h/e^2 \approx 26$ kΩ) as the thickness is reduced, due to enhanced surface scattering and other disorder effects[45],[53]. Altogether, this comparison clearly shows that the WAL 1-to-0 transition observed in the low mobility samples whose surface Fermi levels are far from the surface-hybridization gap at the Dirac point is due to enhanced disorder in the ultrathin regime and not due to the gap opening at the Dirac point.

**5. Conclusion**

Transport measurements on TI materials are often more complex than non-TI materials due to the various topological and non-topological transport channels that are simultaneously present. To address these issues, we have presented a thorough analysis that takes into account the effects of TSS and non-topological bulk channels. By applying the simple Mott and Ioffe-Regel criteria that govern metal-to-insulator transitions, we have shown that it may be thermodynamically impossible to achieve a bulk insulating TI at finite temperatures, because the Mott criterion predicts that such crystals are required to have a bulk defect density below $\sim 10^{14}$ /cm$^3$ (~1 part per billion). However, if upward band-bending effects can be stabilized in these materials and the samples are made thinner than twice the depletion length, it may be possible to overcome the Mott criterion and achieve a true bulk insulating TI. The upward band bending seems to be also essential for clear observation of TSS SdH oscillations. Lastly, we have shown that the number of channels that appear in the WAL effect provides significant evidence pertaining to the electronic properties of the bulk as well as the surface states. It can be either one, two or zero depending on the thickness of the sample, metallicity of the bulk states, location of the surface Fermi level, and the level of disorder. Specifically, the WAL channel can transition from one to zero in ultrathin TI films not only when the Fermi level tunes into the hybridization gap at the Dirac point but also when disorder-driven metal-to-insulator transition occurs. Overall, with this comprehensive examination of the transport properties of the current TI materials, we have cleared up some misunderstandings in the literature and demonstrated that it is possible to achieve a true bulk insulating TI, which will hopefully stimulate and guide future studies in the field of TIs.

**Acknowledgements**

This work is supported by IAMDN of Rutgers University, National Science Foundation (NSF DMR-0845464) and Office of Naval Research (ONR N000141210456).

**Figure captions**

**Fig. 1.** (Color online) The calculated Fermi energy, $E_F$, relative to the conduction band minimum, as a function of buk carrier density. Below a bulk defect density of ~$10^{17}$ /cm$^3$, $E_F$ is within a few meV of the conduction minumum. However, based on the Mott criterion, $E_F$ will stay pinned in the bottom of the conduction band until the bulk defect density drops below the critical value of ~$3 \times 10^{14}$ /cm$^3$.

**Fig. 2.** (Color online) The Mott and Ioffe-Regel criteria showing when a metal-to-insulator transition will occur based on the mobility and the bulk carrier density. The vertical line corresponds to a critical bulk defect density of $N_M \approx 3 \times 10^{14}$ /cm$^3$, while the diagonal lines correspond to $k_Fl$ ~ 0.3-3: carrier densities and mobilities of representative Bi$_2$Se$_3$ and Bi$_2$Te$_3$ family of materials are shown together for comparison. What can be seen from this plot is that no known TI samples are truly an insulator in the Mott sense, and at best they are bad metals, exhibiting weakly insulating temperature dependence. Data are obtained from Ref. [14, 18-29].

**Fig. 3.** (Color online) Band bending direction depending on the surface carrier density. (a) If the surface $E_F$ is exactly at the conduction band minimum, then the surface carrier density for one TSS is equal to ~$5.0 \times 10^{12}$ /cm$^2$ and there is no band bending. (b) If the surface carrier density is above ~$5.0 \times 10^{12}$ /cm$^2$, then the bands must bend downward, and there is an accumulation of carriers at the surface. (c) If the surface carrier density is below ~$5.0 \times 10^{12}$ /cm$^2$, then the bands bend upward, and there is a depletion of bulk carriers near the surface.

**Fig. 4.** (Color online) Band bending parameters for upward band bending. (a) A schematic of the band bending parameters for the case of upward band bending in the infinite thickness limit. $\Delta V$ gives the magnitude of the band bending, $z_D$ is the depletion length, and $E_{DP}$ is the energy seperation of conduction band minimum and the Dirac point. (b) Charge distribution and Fermi levels before equilibrium, determined by local native dopants. (c) Charge distribution and Fermi levels after equilibrium is reached. The charge transfer is given by $n_{CT} = N_{BD} \times z_D$, which causes a potential build-up, and therefore the bands

bend. The Fermi levels are equilibrated everywhere due to the combined effect of the charge transfer, and the shift in potential. This can be made quantative in (d-g) by solving the Poisson equation which determines $\Delta V$, $z_D$, $n_{SS}$, and $n_{CT}$ as a function of intrinsic defect levels of the surface, $n_{SD}$ (vertical axes), and bulk, $N_{BD}$ (horizontal axes).

**Fig. 5.** (Color online) Upward band bending vs thickness. (a) In the bulk limit when the thickness of the TI, $t$, is much larger than the depletion length, $z_D$, the Fermi level sits at the bottom of the conduction band. (b) Band bending in the intermediate thickness ($t \approx 2z_D$). (c) Once the thickness is reduced below $2z_D$, the Fermi level drops below the bottom of the conduction band by an amount $\Delta(t)$. (d) This shows how $\Delta(t)$ changes for various values of the bulk and surface defect densities according to Eq. (10). The dotted lines are for $N_{BD} = 10^{16}$ /cm$^3$, and the solid lines are for $N_{BD} = 10^{17}$ /cm$^3$, and at $t = 0$ nm the upper two curves are for $n_{SD} = 0$ /cm$^2$, the next two are for $n_{SD} = 10^{11}$ /cm$^2$ and the lowest two are for $n_{SD} = 10^{12}$ /cm$^2$. (e-g) shows how $\Delta(t)$ changes for various $N_{BD}$, and $n_{SD}$ while the thickness is fixed. (e) shows $\Delta(t)$ in the ultra thin regime ($t = 10$ nm), where it remains finite over the entire range of $N_{BD}$ and $n_{SD}$ considered. On the other hand, in the bulk limit shown in (g) at $t = 1000$ nm, $\Delta(t)$ is zero over most of the range of $N_{BD}$ and $n_{SD}$.

**Fig. 6.** (Color online) Cartoons showing how weak localizatoin (a) and weak anti-localization (b) occur due to coherenet backscattering along time reversable paths in the absence and presence of the spin-momentum locking, respectively.

**Fig. 7.** (Color online) $\tilde{A}$ parameter in the HLN fomula vs Bi$_2$Se$_3$ film thickness for various set of samples with different qualities. Mobilities and carrier densities of 2, 3 or 6 QL thick sample of each set are shown for quality comparison. (a) High mobility, bulk metallic Bi$_2$Se$_3$ films from Ref. [27] and [52] showing that $\tilde{A}$ is ~1 over the entire thickness range without any 1-to-0 transition down to thickness of 2 QL. (b) High mobility, bulk insulating Cu-doped Bi$_2$Se$_3$ films from Ref. [30] showing that $\tilde{A}$ is ~1 for 2-10 QL and transitions to ~2 for films thicker than ~20 QL: note absence of the 1-to-0 transition for samples down to 2 QL. (c-e) For lower mobility bulk metallic $Bi_2Se_3$ films grown on $Al_2O_3$ (c, Ref. [47]), Si(111) (d, Ref. [45]), and amorphous $SiO_2$ (e), $\tilde{A}$ is nearly constant at ~1 for thick films, but shows a transition from 1 to 0 with decreasing thickness. Note that the WAL 1-to-0 transition strongly depends on the mobilities of the samples. This proves that the metal-insulator transition and the associated WAL 1-to-0 transition observed when the surface Fermi level is far from the Dirac point (c-e) is due to enhanced disorder effect in the ultrathin regime and not due to a surface-hybridization gap opening at the Dirac point.

**Figures**

**Fig. 1. (single-column)**

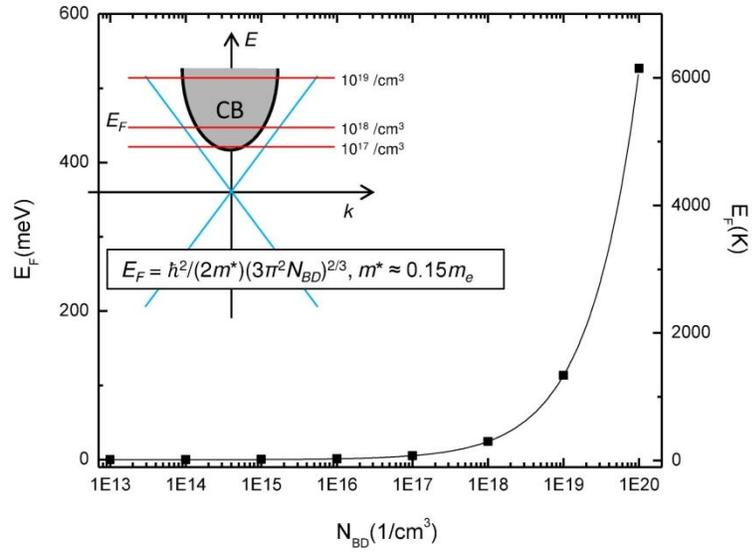

**Fig. 2. (two-column)**

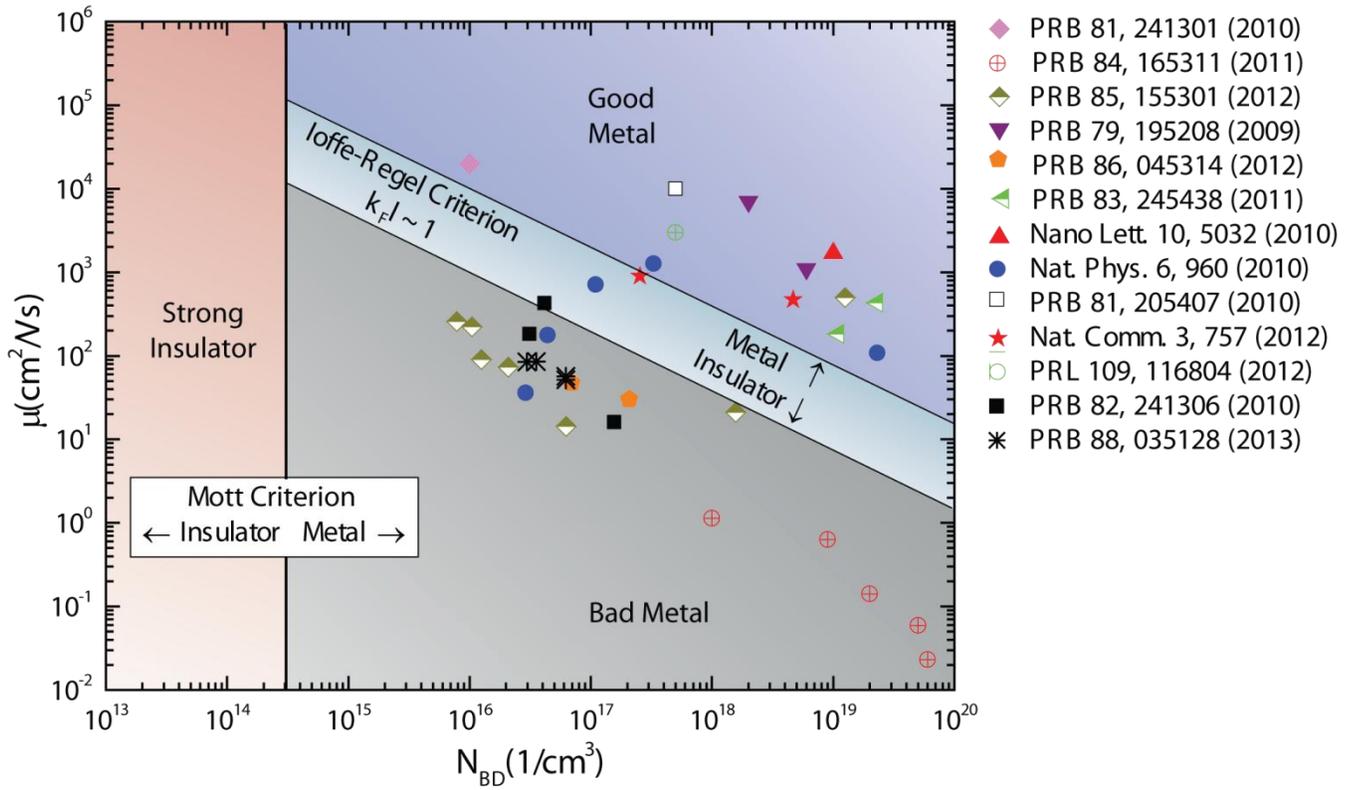

**Fig. 3. (two-column)**

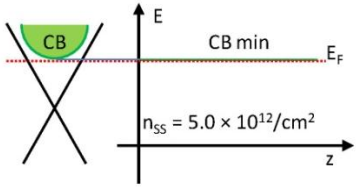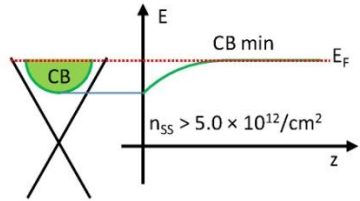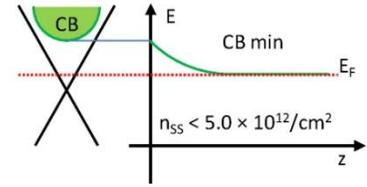

**Fig. 4. (two-column)**

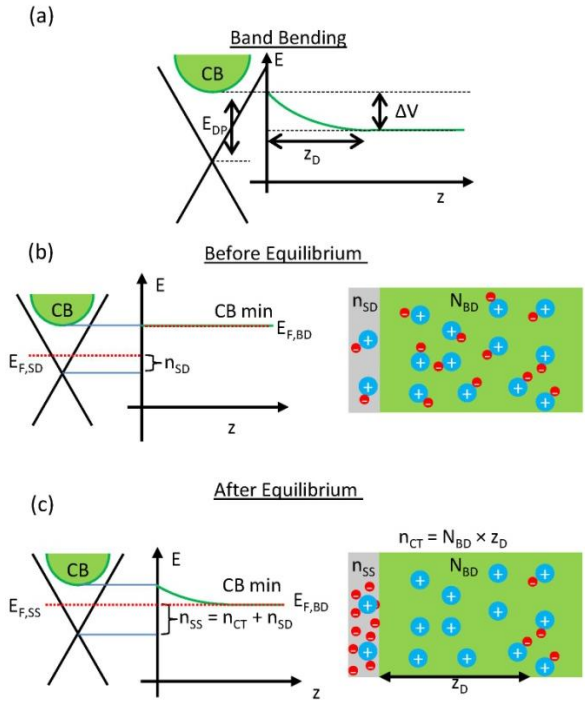
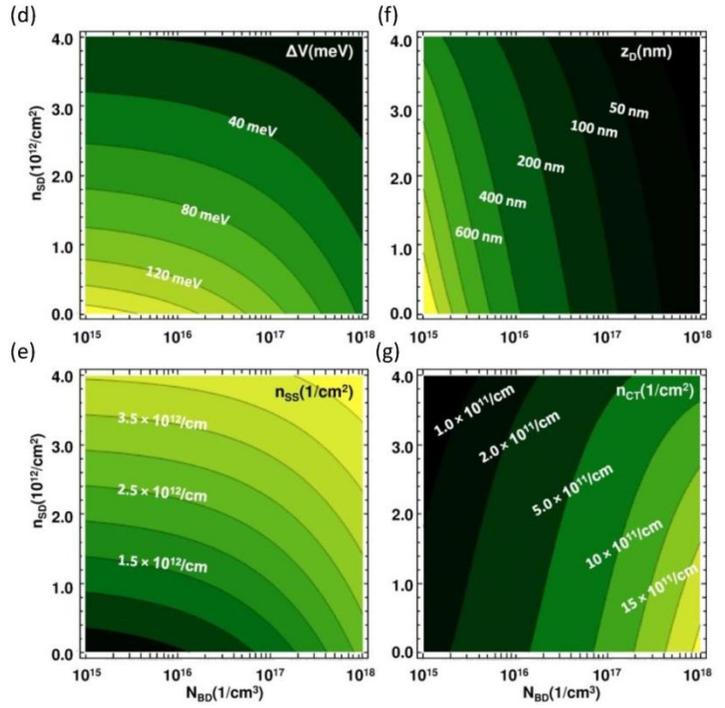

**Fig. 5. (two-column)**

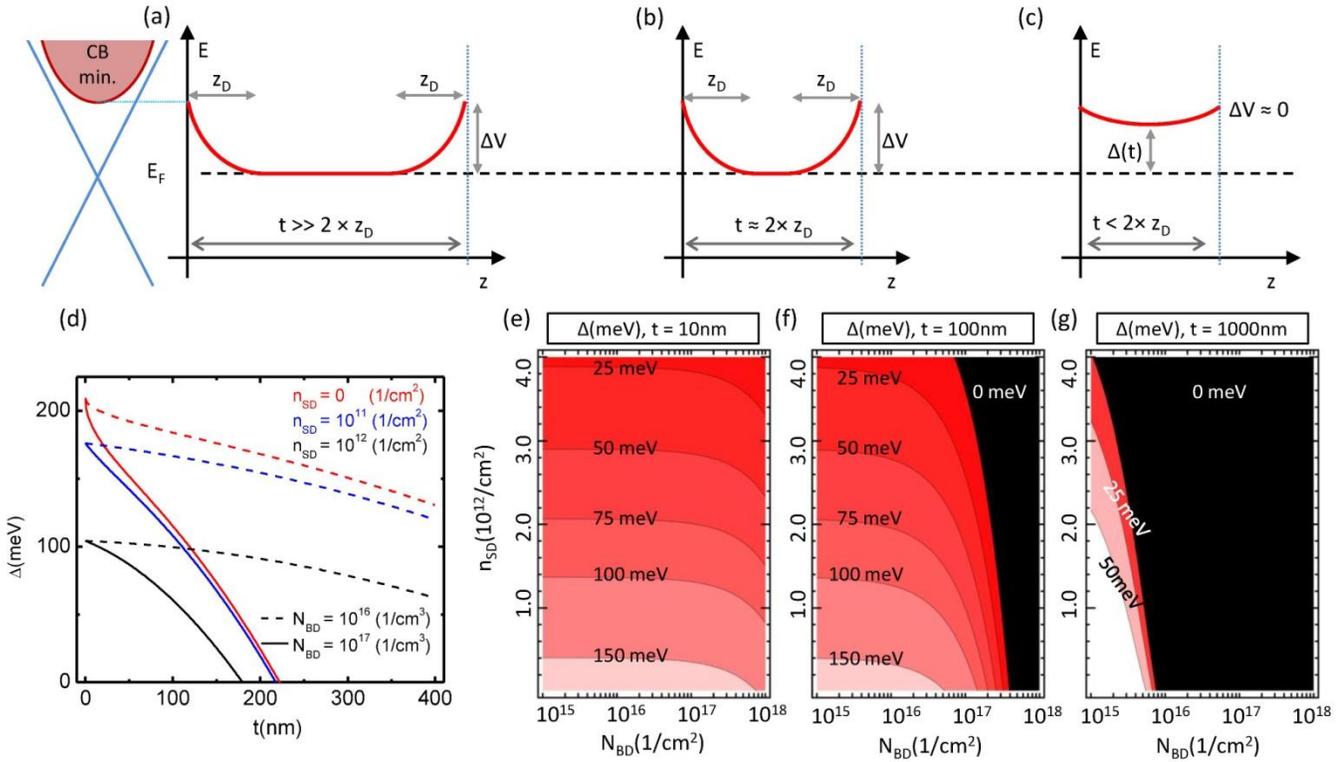

**Fig. 6. (single-column)**

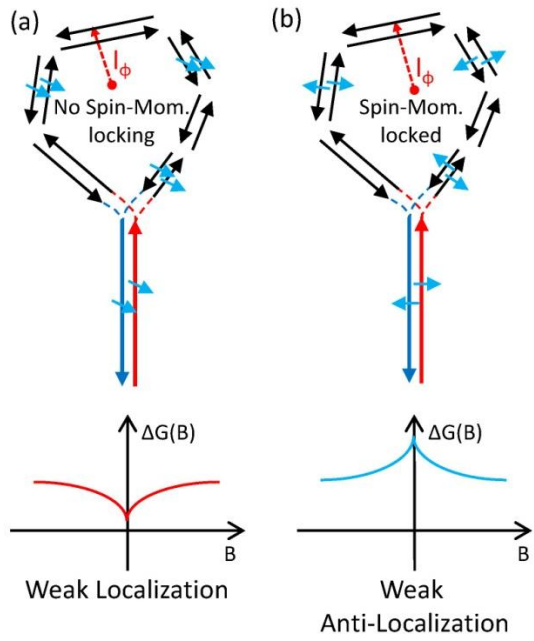

**Fig. 7. (two-column)**

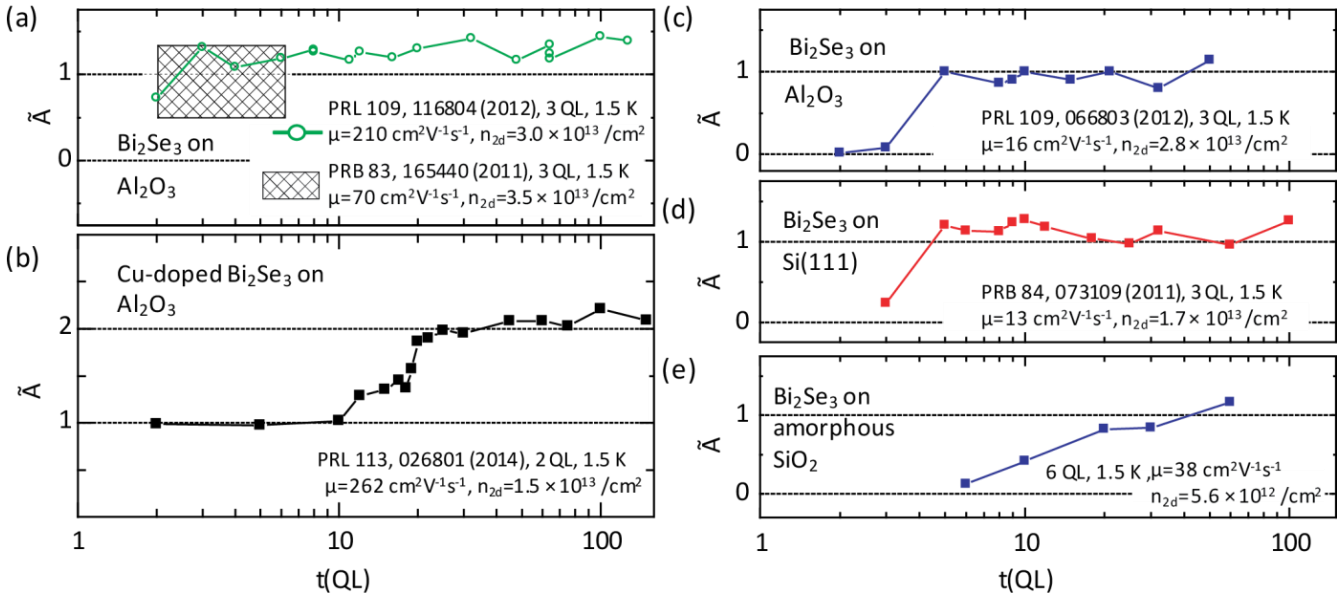